\documentclass[aps,prb,reprint]{revtex4-1}
\usepackage{graphicx}
\usepackage{longtable}
\usepackage{amsmath}
\usepackage{hyperref}
\hypersetup{colorlinks,citecolor=blue,filecolor=black,linkcolor=blue,urlcolor=blue}

\begin{document}
	
	\title{Zn vacancy-donor impurity complexes in ZnO}
	
	\author{Y. K. Frodason}\email[E-mail: ]{ymirkf@fys.uio.no}
	\author{K. M. Johansen}
	\author{T. S. Bj\o rheim}
	\author{B. G. Svensson}
	\affiliation{University of Oslo, Centre for Materials Science and Nanotechnology, N-0318 Oslo, Norway}
	\author{A. Alkauskas}
	\affiliation{Center for Physical Sciences and Technology, Vilnius LT-10257, Lithuania}	
	
	\date{\today}
	
\begin{abstract}

Results from hybrid density functional theory calculations on the thermodynamic stability and optical properties of the Zn vacancy ($V_{\text{Zn}}$) complexed with common donor impurities in ZnO are reported. Complexing $V_{\text{Zn}}$ with donors successively removes its charge-state transition levels in the band gap, starting from the most negative one. Interestingly, the presence of a donor leads only to modest shifts in the positions of the $V_{\text{Zn}}$ charge-state transition levels, the sign and magnitude of which can be interpreted from a polaron energetics model by taking hole-donor repulsion into account. By employing a one-dimensional configuration coordinate model, luminescence lineshapes and positions were calculated. Due to the aforementioned effects, the isolated $V_{\text{Zn}}$ gradually changes from a mainly non-radiative defect with transitions in the infrared region in \textit{n}-type material, to a radiative one with broad emission in the visible range when complexed with shallow donors.

\end{abstract}
	
\maketitle	

\section{\label{sec:introduction}Introduction}
	
Zinc oxide---a II-VI compound semiconductor with a wide direct band gap of 3.44 eV \cite{Reynolds1999}---has been studied quite extensively over the past decade. Its unique optical and electrical properties, including an exceptionally large free exciton binding energy of 60 meV, make it attractive for optoelectronic devices. An ongoing issue holding back this development, however, is the lack of stable and reproducible \textit{p}-type ZnO material. A prerequisite for a successful application of any semiconductor is a good understanding of native defects, common impurities, dopants and their interplay. To this end, unambiguous spectroscopic identification of key defects is essential.
	
Using photoluminescence (PL) spectroscopy, several broad luminescence bands peaking in the visible part of the emission spectrum have been observed, both in as-grown and processed ZnO samples. These bands are known to originate from deep level centers, but very few have been unambiguously identified. One example is the characteristic red luminescence band, hereby referred to as the RL\footnote{This RL should be distinguished from the RL at $\sim$1.7 eV, attributed to the $\text{N}_{\text{O}}$ acceptor in ZnO \cite{Tarun2011}.}, peaking at about 1.8 eV and exhibiting a low activation energy of thermal quenching of 10--20 meV \cite{Vlasenko2005,Kappers2008,Evans2008,Knutsen2012,Wang2009,Reshchikov2006,Reshchikov2007}. The RL has been the subject of numerous experimental studies as its intensity can be raised significantly via high-energy electron irradiation at room temperature (RT) \cite{Vlasenko2005,Kappers2008,Evans2008,Knutsen2012}. While a number of native defects have been invoked as candidates, a large body of evidence has accumulated towards a donor-acceptor pair (DAP) transition involving the $V_{\text{Zn}}$ acceptor and a residual shallow donor \cite{Kappers2008,Evans2008,Knutsen2012,Wang2009}. However, the exact configuration and nature of the defect responsible for the RL remains elusive.
	
$V_{\text{Zn}}$ is the dominant native acceptor-type defect in ZnO \cite{Tuomisto2003}, acting as a compensating center in \textit{n}-type material. Moreover, $V_{\text{Zn}}$ can trap a localized hole polaron on each of its four nearest-neighbor O ions \cite{Lany2009,Frodason2017}. Previously, we have shown that subsequent electron capture by $V_{\text{Zn}}$ can give rise to broad luminescence bands in the infrared (IR) part of the emission spectrum in \textit{n}-type material, although the transitions are predominantly most likely nonradiative \cite{Frodason2017}. In the present work, we investigate the thermodynamic and optical properties of $V_{\text{Zn}}$ complexed with shallow donor impurities that are commonly present in as-grown ZnO crystals, namely H, Al, Ga and Si. Specifically, we have studied ($V_{\text{Zn}}\text{Al}_{\text{Zn}}$), ($V_{\text{Zn}}\text{Ga}_{\text{Zn}}$), ($V_{\text{Zn}}\text{Si}_{\text{Zn}}$) and ($V_{\text{Zn}}n\text{H})$ with \textit{n}=1,2,3, as well as three complexes with a combination of different donors, namely ($V_{\text{Zn}}\text{Al}_{\text{Zn}}$H), ($V_{\text{Zn}}\text{Al}_{\text{Zn}}$2H) and ($V_{\text{Zn}}\text{Si}_{\text{Zn}}$H). Many of these $V_{\text{Zn}}$-donor complexes have already been identified in processed ZnO samples by electron paramagnetic resonance (EPR) \cite{Kappers2008,Evans2008,Son2014,Stehr2014,Holston2016} and infrared (IR) spectroscopy \cite{Lavrov2002,Lavrov2007,Bastin2011,Herklotz2010,Herklotz2015}. Our theoretical predictions are consistent with such experimental data, and, importantly, provide a guide for future studies of the broad luminescence bands observed in ZnO, and identification of $V_{\text{Zn}}$-donor complexes.
		
\section{\label{sec:methodology}Methodology}
	
A comprehensive description of the employed first-principles methodology and computational details can be found in Ref. \onlinecite{Frodason2017}. A brief summary is given here.
	
Unless specified, all calculations were based on the generalized Kohn-Sham theory with the Heyd-Scuseria-Ernzerhof (HSE) \cite{Krukau2006} hybrid functional and the projector augmented wave method \cite{Bloechl1994,Kresse1994,Kresse1999}, as implemented in the VASP code \cite{Kresse1993,Kresse1996}. The amount of screened Hartree-Fock exchange was set to $\alpha = 37.5 \%$, giving band gap (3.42 eV) and lattice parameters ($a=3.244$ \AA \ and $c=5.194$ \AA) in excellent agreement with experimental data \cite{Reynolds1999,Albertsson1989}. Defect calculations were performed using a 96-atom-sized supercell, a plane-wave energy cutoff of 500 eV, and a special off-$\Gamma$ \textit{k}-point at $k=(\frac{1}{4},\frac{1}{4},\frac{1}{4})$ \cite{Baldereschi1973}. Ionic relaxation was carried out consistently with the hybrid functional, and spin-polarization was explicitly included. 
	
Defect formation energies and thermodynamic charge-state transition levels were determined by following the standard formalism \cite{Zhang1991,Freysoldt2014}. The chemical potential of Ga, Al, Si and H were referenced to Ga$_{2}$O$_{3}$(s), Al$_{2}$O$_{3}$(s), SiO$_{2}$(s) and H$_{2}$O(g)/H$_{2}$(g), respectively.  Corrections for the spurious long-range Coulomb interactions between charged defects, their periodic images and the neutralizing jellium were included by employing the extended \cite{Kumagai2014} Freysoldt-Neugebauer-Van de Walle scheme \cite{Freysoldt2009,Komsa2012}.
	
\begin{figure}[!htb]
	\includegraphics[width=\columnwidth]{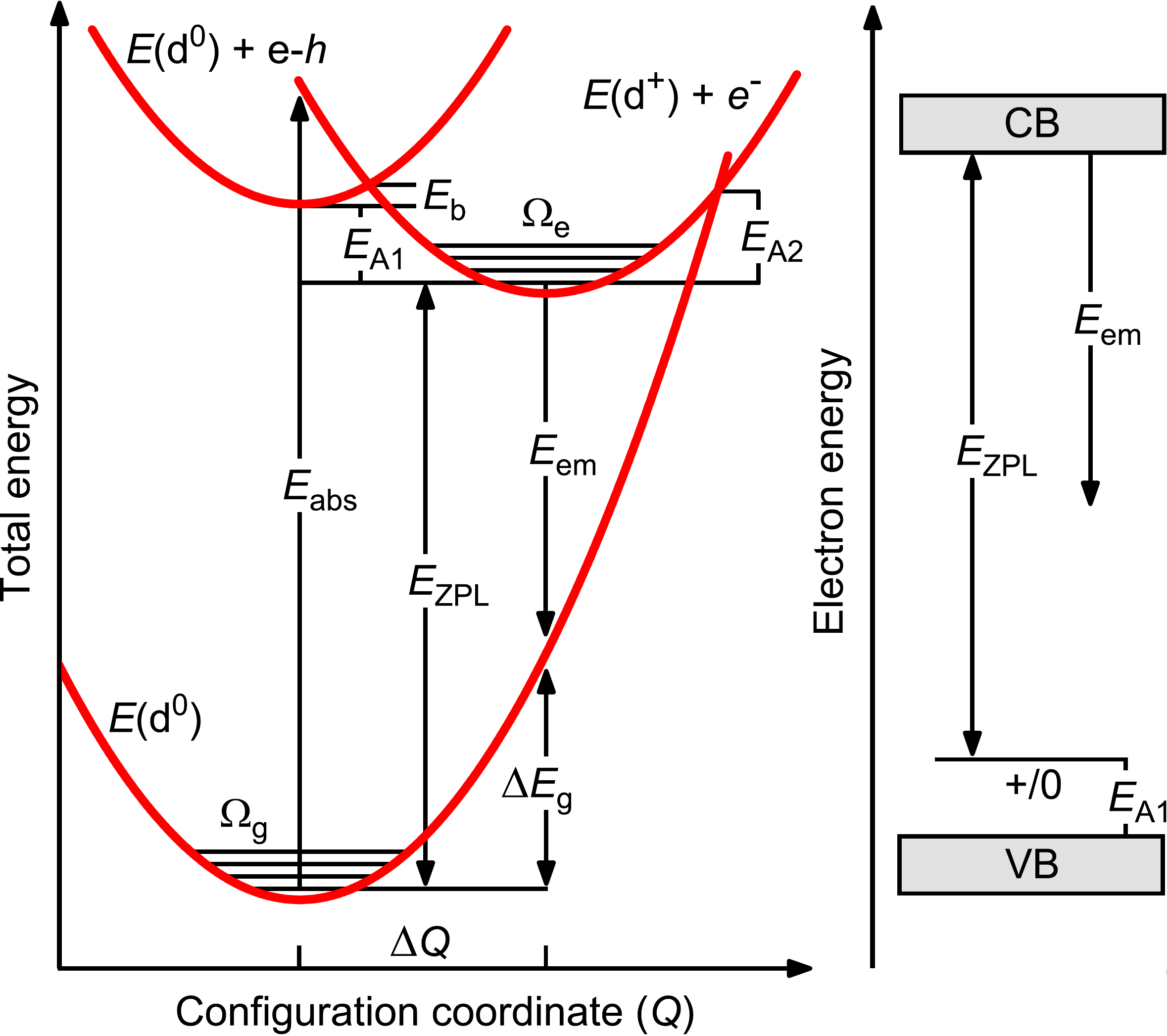}
	\caption{A configuration coordinate diagram is shown on the left side, illustrating vibronic transitions between the ground (neutral) and excited (singly positive) charge-state of a defect. Free electrons (\textit{e}) and holes (\textit{h}) are located at the conduction band (CB) minimum and valence band (VB) maximum, respectively. $E_{\text{ZPL}}$ gives the thermodynamic charge-state transition level relative to the CB minimum (shown in the band diagram on the right). $E_{\text{b}}$ is the barrier for capture of a free hole by the neutral defect, $E_{\text{A}1}$ gives the thermodynamic charge-state transition level relative to the VB maximum, i.e., the energy required for thermal emission of a hole from the defect to the VB, and $E_{\text{A}2}$ is the barrier for nonradiative capture of an electron at the CB minimum \cite{Reshchikov2014a}.
	}
	\label{fig:cc_diagram}
\end{figure}

As in Ref. \onlinecite{Frodason2017}, defect luminescence lineshapes and positions were calculated by utilizing the effective one-dimensional configuration coordinate (CC) model \cite{Alkauskas2012,Stoneham}, shown in Fig. \ref{fig:cc_diagram}, where the configuration coordinate $Q$ connects the initial and final states. The parameters that enter the model are the effective phonon frequencies $\Omega_{\text{g/e}}$, the zero phonon line (ZPL) energy $E_{\text{ZPL}}$ and the change in configuration coordinate $\Delta\text{Q}$, all of which are obtained from the hybrid density functional theory calculations. Within the classical Franck Condon approximation, emission (em) and absorption (abs) energies may be defined as $E_{\text{em}} = E_{\text{ZPL}} - \Delta E_{\text{g}}$ and $E_{\text{abs}} = E_{\text{ZPL}} + \Delta E_{\text{e}}$, respectively. Here, $\Delta E_{\text{g}}$ and $\Delta E_{\text{e}}$ are the ground (g) and excited (e) state relaxation energies, often referred to as Franck-Condon shifts. Finally, Huang-Rhys (HR) factors \cite{Huang1950,Stoneham} describe the average number of phonons that are involved in a transition, and can be expressed as $S_{\text{g/e}} = \Delta E_{\text{g/e}}/\hbar\Omega_{\text{g/e}}$ for emission/absorption.  The effective one-dimensional CC model is a good approximation for defects with strong electron-phonon coupling (large HR factors), as numerically shown in Refs. \onlinecite{Alkauskas2012,Alkauskas2016}, and this is indeed the case for $V_{\text{Zn}}$ in ZnO \cite{Frodason2017,Alkauskas2012}.

\section{\label{sec:results_and_discussion}Results and Discussion}

\subsection{\label{sec:defect_configuration}Defect complex configuration}

\begin{figure*}[!htb]
	\includegraphics[width=0.95\textwidth]{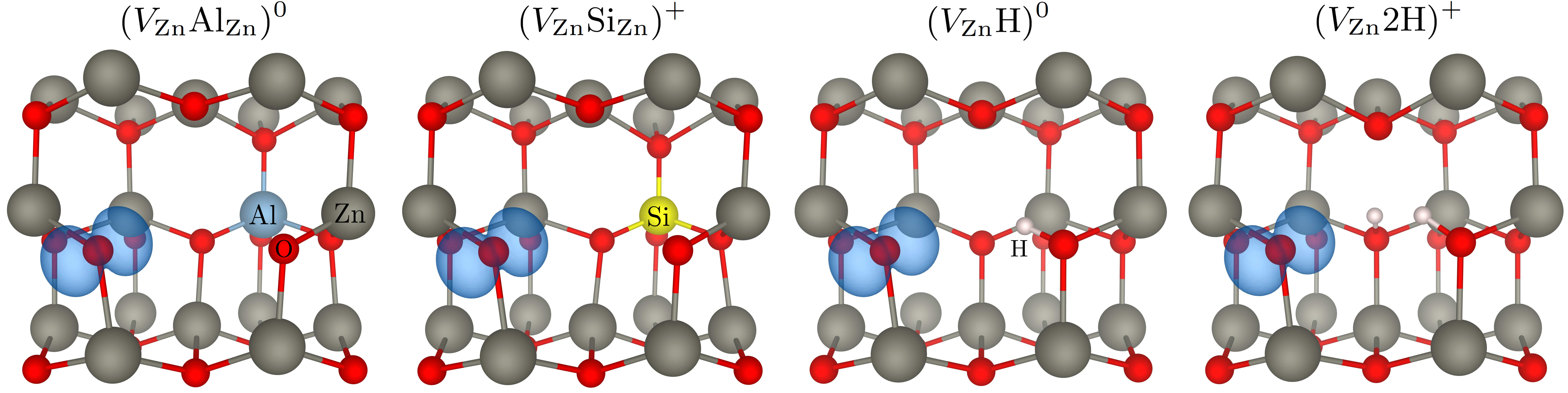}
	\caption{Four representative $V_{\text{Zn}}$-donor complexes are shown with a single hole trapped in a polaronic O 2\textit{p}-like state (blue isosurface at 0.01 $r_{\text{Bohr}}^{-3}$). Subsequent capture of an electron at the CB can potentially give rise to broad luminescence.}
	\label{fig:configuration}
\end{figure*}

$V_{\text{Zn}}$ is tetrahedrally surrounded with a dangling O 2\textit{p} bond at each corner. In the neutral charge-state, $V_{\text{Zn}}^{0}$ has two holes, which occupy polaronic states on the nearest-neighbor O ions, and can be filled by electrons. By forming a complex with $V_{\text{Zn}}$, a donor may directly supply one or both of these electrons. Thus, e.g., $(V_{\text{Zn}}2\text{H})^{0}$ will be neutral with all defect levels filled. As shown in Fig. \ref{fig:configuration}, such complexes can still trap holes in polaronic states.

Donors can bind to $V_{\text{Zn}}$ in several different configurations. Al, Ga and Si substitute on a neighboring Zn site, and can either share the same Zn (001) plane as $V_{\text{Zn}}$, or reside in one of the two nearest neighbor Zn (001) planes. As shown for $(V_{\text{Zn}}\text{Al}_{\text{Zn}})$ and $(V_{\text{Zn}}\text{Si}_{\text{Zn}})$ in Fig. \ref{fig:configuration}, we have explored the shared Zn-plane configuration, since this is the lowest energy configuration determined by the generalized gradient approximation in the Perdew-Burke-Ernzerhof \cite{Perdew1996} parametrization. Our calculations indicate that hole localization occurs on the O ion furthest away from the positively charged donor, as also found experimentally for $(V_{\text{Zn}}\text{Al}_{\text{Zn}})^{0}$ using EPR \cite{Stehr2014}. 

H can bind to $V_{\text{Zn}}$ by terminating one of the four dangling O 2\textit{p} bonds \cite{Lavrov2002}. Both axial and azimuthal O--H bond configurations of ($V_{\text{Zn}}n\text{H}$) complexes with \textit{n}=1,2,3 have been assigned to IR local vibrational modes \cite{Lavrov2002,Lavrov2007,Bastin2011,Herklotz2010,Herklotz2015}. We obtain very small differences in total energy between these configurations for the ($V_{\text{Zn}}\text{H}$), ($V_{\text{Zn}}2\text{H}$) and ($V_{\text{Zn}}3\text{H}$) complexes. Hence, both configurations are expected to occur, in agreement with IR spectroscopy data \cite{Herklotz2010,Herklotz2015}. In principle, $(V_{\text{Zn}}3\text{H})$ can capture a fourth H, but we find $(V_{\text{Zn}}4\text{H})^{2+}$ to be unstable with respect to $(V_{\text{Zn}}3\text{H})^{+}$ and $\text{H}_{i}^{+}$. Moreover, its formation will likely be suppressed by Coulomb repulsion as both constituents are positively charged in the entire range of Fermi level positions in the band gap \cite{Hupfer2017,Walle2000}. This logic holds for other singly positively charged complexes as well, i.e, $(V_{\text{Zn}}\text{Al}_{\text{Zn}}\text{2H})^{+}$ and $(V_{\text{Zn}}\text{Si}_{\text{Zn}}\text{H})^{+}$ are unlikely to accept one more donor. These complexes are considered as saturated with donors.

\subsection{\label{sec:defect_energetics}Defect complex charge-state transition levels}

Fig. \ref{fig:formation_energy} shows the formation energy of the isolated $V_{\text{Zn}}$ and the $V_{\text{Zn}}$-donor complexes as a function of the Fermi level position in the O-rich limit. Before addressing the complexes, we note that the results for the isolated $V_{\text{Zn}}$ scatter widely in the literature with regard to the position and number of thermodynamic charge-state transition levels. The main reasons for these variations were discussed previously in Ref. \onlinecite{Frodason2017}. Furthermore, the $(V_{\text{Zn}}\text{Ga}_{\text{Zn}})$ complex is omitted from further discussion, as its charge-state transition levels are identical to those of the $(V_{\text{Zn}}\text{Al}_{\text{Zn}})$ complex (albeit with a 0.2 eV lower formation energy). Regarding the positions of the charge-state transition levels in Fig. \ref{fig:formation_energy}, there are two clear trends: 

\begin{figure}[htb]
	\includegraphics[width=\columnwidth]{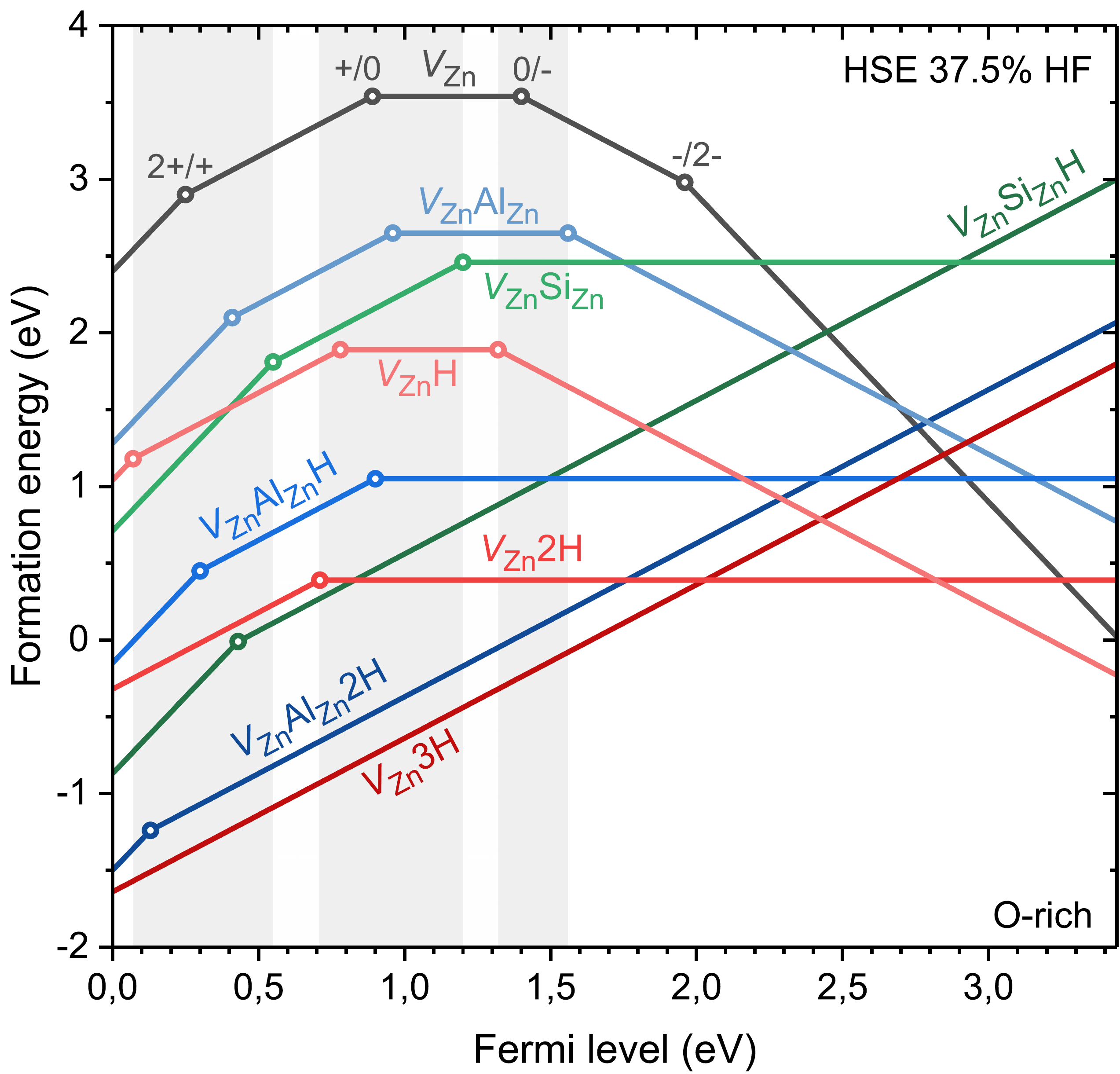} 
	\caption{Formation energy of $V_{\text{Zn}}$-donor complexes as a function of the Fermi level position from the VB maximum to the CB minimum, under O-rich conditions. The shaded vertical bars in the left panel indicate the Fermi level range of the charge-state transition levels of $V_{\text{Zn}}$ and the complexes.} \label{fig:formation_energy}
\end{figure}

(i) Complexing $V_{\text{Zn}}$ with donors successively removes its charge-state transition levels in the band gap, starting from the most negative one. Al$_{\text{Zn}}$, Ga$_{\text{Zn}}$ and H are single donors and remove one transition level, while Si$_{\text{Zn}}$ is a double donor and removes two transition levels. If the number of donor electrons exceeds two, the complex becomes a donor with characteristics similar to those of the isolated donor. For instance, ($V_{\text{Zn}}3\text{H}$) becomes a shallow single donor where the third donor electron occupies a hydrogenic effective mass state just below the CB \cite{Walle2000}.

(ii) Similar to $V_{\text{Ga}}$-donor complexes in GaN \cite{Lyons2015}, the thermodynamic charge-state transition levels of the isolated $V_{\text{Zn}}$ are not overly affected by the presence of donors; the respective levels shift within narrow ranges in the band gap, as indicated in Fig. \ref{fig:formation_energy}. Interestingly, the transition levels of complexes containing H shift down, while those containing other donors shift up, relative to the isolated $V_{\text{Zn}}$ levels. This behavior can be interpreted through the polaron energetics model introduced in Ref. \onlinecite{Frodason2017}, by also considering the electrostatic repulsion between hole polarons and donors. For example, consider the (0/-) transition levels. Adding a hole polaron to $V_{\text{Zn}}^{-}$ lowers its energy by the hole polaron addition energy $\varepsilon_{0}$ minus the hole-hole repulsion $U_{\text{hole}}$. When adding a hole polaron to the $(V_{\text{Zn}}\text{Al}_{\text{Zn}})^{-}$ complex, however, $U_{\text{hole}}$ is replaced by the hole-$\text{Al}^{+}_{\text{Zn}}$ repulsion $U_{\text{Al}}$. Since $\text{Al}^{+}_{\text{Zn}}$ is located outside the vacancy, $U_{\text{Al}}<U_{\text{hole}}$, and thus the (0/-) transition level of $(V_{\text{Zn}}\text{Al}_{\text{Zn}})$ shifts up in energy, relative to the isolated $V_{\text{Zn}}$. Similar arguments apply to the ($V_{\text{Zn}}\text{H})$ complex; since H$^{+}$ resides within the vacancy, it causes the O$^{2-}$ tetrahedron surrounding the vacancy to contract. As a result, the hole-H$^{+}$ distance in ($V_{\text{Zn}}\text{H})^{0}$ is significantly shorter than the hole-hole distance in $V_{\text{Zn}}^{0}$, and so $U_{\text{H}}>U_{\text{hole}}$, and the (0/-) level shifts down in energy. Estimating the hole-donor repulsion, the polaron energetics model can be used to qualitatively predict the energy position of the charge-state transition levels of any $V_{\text{Zn}}$-donor complex, relative to those of $V_{\text{Zn}}$.  

For ($V_{\text{Zn}}3\text{H})$, we find that the hole-donor repulsion is too strong for a hole polaron to become trapped, i.e., ($V_{\text{Zn}}3\text{H})^{2+}$ is unstable. However, based on the polaron energetics model, the hole-donor repulsion is anticipated to be smaller if one or two H$^{+}$ ions in ($V_{\text{Zn}}3\text{H})$ are replaced by $\text{Al}^{+}_{\text{Zn}}$, $\text{Ga}^{+}_{\text{Zn}}$ or $\text{Si}^{2+}_{\text{Zn}}$. For this reason, we have investigated the $(V_{\text{Zn}}\text{Al}_{\text{Zn}}2\text{H})$ and $(V_{\text{Zn}}\text{Si}_{\text{Zn}}\text{H})$ complexes. Interestingly, we find that a hole polaron can be stabilized for $(V_{\text{Zn}}\text{Al}_{\text{Zn}}2\text{H})$, with the (2+/+) level occurring 0.13 eV above the VB maximum. Similarly, when replacing two H$^{+}$ ions in ($V_{\text{Zn}}3\text{H})$ with $\text{Si}^{2+}_{\text{Zn}}$, the (2+/+) level of the resulting $(V_{\text{Zn}}\text{Si}_{\text{Zn}}\text{H})$ complex is located even deeper in the gap, at 0.43 eV above the VB maximum. To complete the picture, $(V_{\text{Zn}}\text{Al}_{\text{Zn}}\text{H})$ was also investigated. The (+/0) level of this complex occurs 0.90 eV above the VB maximum, approximately halfway between the (+/0) level of $(V_{\text{Zn}}\text{Al}_{\text{Zn}})$ and $(V_{\text{Zn}}\text{H})$ as expected.

\begin{figure}[!htb]
	\includegraphics[width=\columnwidth]{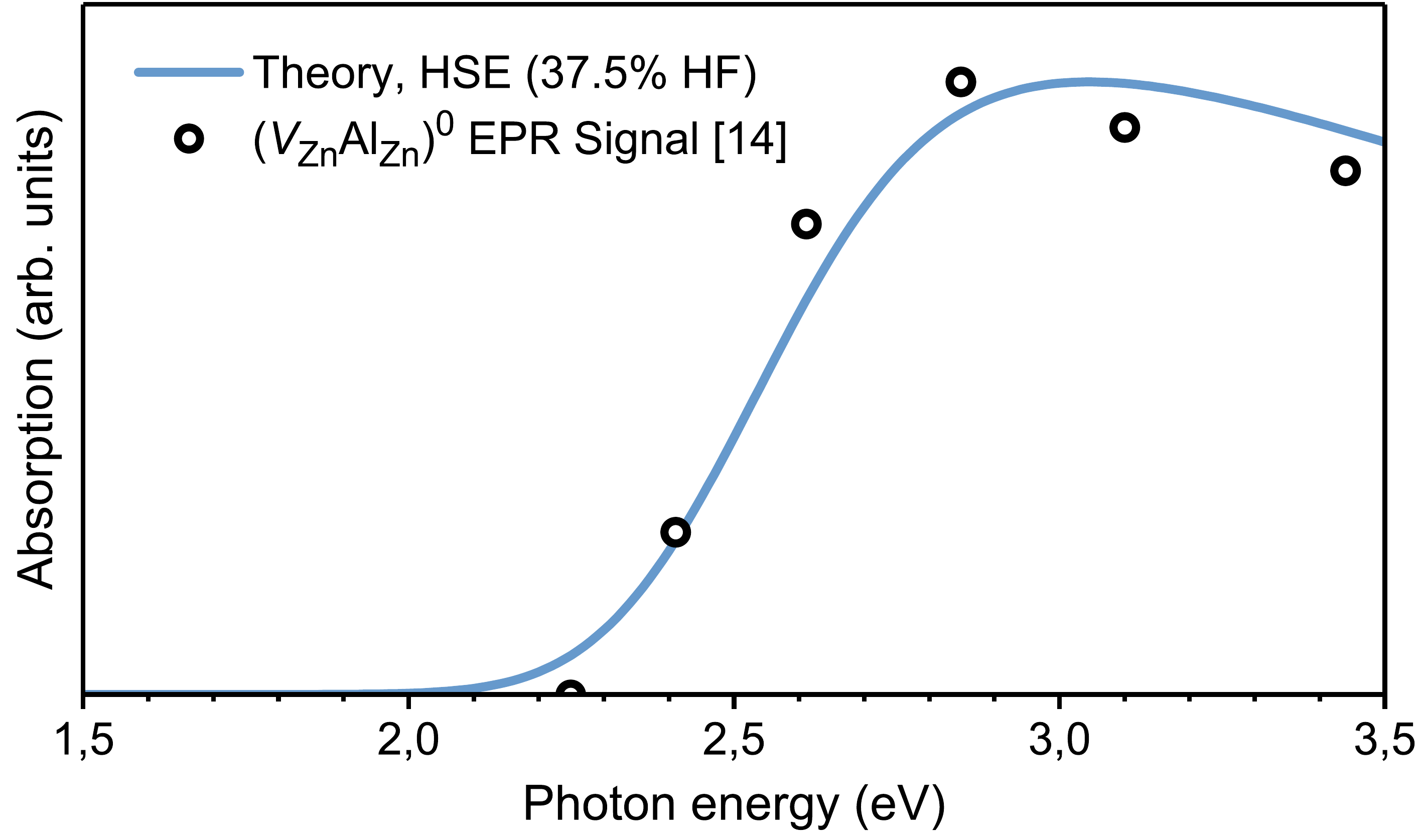}
	\caption{Simulated absorption cross section for $(V_{\text{Zn}}\text{Al}_{\text{Zn}})^{-}$, and EPR signal of $(V_{\text{Zn}}\text{Al}_{\text{Zn}})^{0}$ as a function of the photon energy of light illumination \cite{Stehr2014}. The peak of the calculated profile is normalized to the maximum EPR data point.}
	\label{fig:photo_epr}
\end{figure}

The calculated (0/$-$) transition level of $(V_{\text{Zn}}\text{Al}_{\text{Zn}})$ exhibits close agreement with recent photo-EPR data; Stehr \textit{et al.} \cite{Stehr2014} detected the EPR signal of $(V_{\text{Zn}}\text{Al}_{\text{Zn}})^{0}$ only when the photon energy exceeded $\sim$2.4 eV. Based on the CC model, we obtain a classical absorption energy of 2.54 eV. However, the absorption onset will be lower due to vibrational broadening \cite{Alkauskas2016}. To enable a more valid comparison, we have simulated the absorption profile, including vibrational broadening, by following the scheme outlined in Ref. \onlinecite{Kopylov1975}. As shown by the relative comparison in Fig. \ref{fig:photo_epr}, the resulting absorption profile is in good agreement with the photo-EPR data \cite{Stehr2014}. Moreover, Stehr \textit{et al.} \cite{Stehr2014a,Stehr2014} detected a signal from the isolated $V_{\text{Zn}}^{-}$ under illumination with photon energies exceeding $\sim$2.1 eV. If one assumes that the vibrational broadening for $(V_{\text{Zn}}\text{Al}_{\text{Zn}})$ and $V_{\text{Zn}}$ is similar, the experimental data suggest a difference of $\sim$0.3 eV between the (0/$-$) level of $(V_{\text{Zn}}\text{Al}_{\text{Zn}})$ and ($-$/$2-$) level of $V_{\text{Zn}}$, which is in good agreement with our calculations (0.40 eV) and consistent with the polaron energetics model ($U_{\text{Al}}<U_{\text{hole}}$). The calculated (0/$-$) transition level of $(V_{\text{Zn}}\text{H})$ similarly exhibits close agreement with photo-EPR data by Evans \textit{et al.} \cite{Evans2008}.

\subsection{\label{sec:}Defect complex thermodynamics}

An important question is whether $V_{\text{Zn}}$-donor complexes are likely to be present in as-grown ZnO samples and not just in post-growth processed ones, and if they are stable at RT. A stable complex implies that the defect reaction to form the complex lowers the total energy, i.e., the defect complex must have a positive ``removal'' energy. The removal energy is defined as the difference in formation energy between the defect complex and the isolated constituents (in their most stable configuration) when one donor is removed from the complex. For instance, the removal energy of ($V_{\text{Zn}}2\text{H})$ is given by
\begin{equation*}\label{key}
	E^{\text{r}}[(V_{\text{Zn}}2\text{H})^{0}] = E^{\text{f}}[(V_{\text{Zn}}\text{H})^{-}] + E^{\text{f}}(\text{H}_{i}^{+}) - E^{\mathrm{f}}[(V_{\text{Zn}}2\text{H})^{0}],
\end{equation*}
which is well defined, as all terms in the equation pertain to the lowest-energy charge configuration for \textit{n}-type conditions. The removal and formation energy of the $V_{\text{Zn}}$-donor complexes and their constituents, for \textit{n}-type conditions ($\varepsilon_{F}$ at CBM), are provided in Table \ref{table:removal-energy}. In the case of $V_{\text{Zn}}$-donor complexes with a combination of different donors, the energy for removal of H is given. The removal energies of $(V_{\text{Zn}}\text{Al}_{\text{Zn}})$, $(V_{\text{Zn}}\text{Ga}_{\text{Zn}})$, $(V_{\text{Zn}}\text{H})$ and $(V_{\text{Zn}}2\text{H})$ are in good agreement with previous HSE calculations by Steiauf \textit{et al.} \cite{Steiauf2014} and Lyons \textit{et al.} \cite{Lyons2017}.

\begin{table}[!htb]
	\caption{Formation and removal energy of $V_{\text{Zn}}$-donor complexes and their constituents under \textit{n}-type conditions. Formation energies are provided for O-rich, O-poor and intermediate (halfway between O-rich and O-poor) conditions. \label{table:removal-energy}}
	\begin{ruledtabular}
		\begin{tabular}{lcccc}
			 & \multicolumn{3}{c}{$E^{\text{f}}$($\varepsilon_{F}$ at CBM) (eV)} & \\
			Defect & O-rich & intermediate & O-poor & $E^{\text{r}}$ (eV) \\
			\hline
			$\text{H}_{i}^{+}$ & 2.79 & 2.05 & 1.31 & - \\
			$\text{Al}_{\text{Zn}}^{+}$ & 1.99 & 1.12 & 0.25 & - \\
			$\text{Ga}_{\text{Zn}}^{+}$ & 1.76 & 0.89 & 0.02 & - \\
			$\text{Si}_{\text{Zn}}^{2 +}$ & 4.90 & 3.16 & 1.41 & - \\
			$V_{\text{Zn}}^{2-}$ & 0.02 & 1.77 & 3.51 & - \\		
			\hline	
			$(V_{\text{Zn}}\text{H})^{-}$ & -0.23 & 1.01 & 1.78 & 3.04 \\	
			$(V_{\text{Zn}}2\text{H})^{0}$ & 0.39 & 0.67 & 0.92 & 2.17 \\	
			$(V_{\text{Zn}}3\text{H})^{+}$ & 1.80 & 1.33 & 0.85 & 1.38 \\
			$(V_{\text{Zn}}\text{Al}_{\text{Zn}})^{-}$ & 0.77 & 1.65 & 2.52 & 1.24 \\
			$(V_{\text{Zn}}\text{Al}_{\text{Zn}}\text{H})^{0}$ & 1.05 & 1.19 & 1.32 & 2.51 \\
			$(V_{\text{Zn}}\text{Al}_{\text{Zn}}2\text{H})^{+}$ & 2.07 & 1.47 & 0.86 & 1.77 \\
			$(V_{\text{Zn}}\text{Ga}_{\text{Zn}})^{-}$ & 0.57 & 1.45 & 2.32 & 1.21 \\
			$(V_{\text{Zn}}\text{Si}_{\text{Zn}})^{0}$ & 2.46 & 2.46 & 2.46 & 2.46 \\
			$(V_{\text{Zn}}\text{Si}_{\text{Zn}}\text{H})^{+}$ & 3.02 & 2.28 & 1.54 & 2.23 \\
		\end{tabular}
	\end{ruledtabular}
\end{table}

All the removal energies are positive and large, meaning that the complexes are expected to be stable at RT. However, assuming thermodynamic equilibrium at the growth temperature, a positive removal energy does not necessarily mean that a sizable fraction of constituents will form complexes during materials growth \cite{Walle2004}. This is because complex formation generally lowers the configurational entropy. In order for the equilibrium concentration of a defect complex to be larger than that of either constituent, its formation energy should generally be lower than that of both constituents \cite{Freysoldt2014,Walle2004}. Considering the intermediate formation energies given in Table \ref{table:removal-energy} (the O-rich/poor limits are not usually accessed during materials growth), most of the $V_{\text{Zn}}$-donor complexes have a formation energy that is lower than or at least comparable to that of their constituents. Consequently, such $V_{\text{Zn}}$-donor complexes, and especially the $(V_{\text{Zn}}n\text{H})$ ones, are likely to form even during materials growth. Note that the complexes may also form under non-equilibrium conditions. For instance, if the individual constituents are incorporated during growth, complexes can form during cooldown \cite{Walle2004}. In this scenario, formation of complexes involving H is anticipated to take place rapidly, because of the high mobility of $\text{H}_{i}$ even at RT \cite{Wardle2006,Johansen2008,Hupfer2016,Hupfer2017}

The absolute and relative concentration of different $V_{\text{Zn}}$-donor complexes will depend on the synthesis method used, governed by the presence of $V_{\text{Zn}}$ and impurities during the growth. Indeed, according to secondary ion mass spectrometry (SIMS) data \cite{Vines2013}, the concentration of Al and Si in as-grown ZnO single crystals typically varies between 10$^{15}$ and 10$^{17}$ cm$^{-3}$ for different growth techniques (Ga is less abundant \cite{McCluskey2007}), while the concentration of H is below the typical SIMS detection limit of $5\times10^{17}$ cm$^{-3}$. In addition, several kinds of different traps compete with $V_{\text{Zn}}$ for $\text{H}_{i}$, such as $\text{Li}_{\text{Zn}}$ \cite{Johansen2011,Heinhold2017,Herklotz2015}, a common impurity in hydrothermally grown (HT) ZnO.  Moreover, despite $V_{\text{Zn}}$ being the most likely intrinsic defect to form under realistic growth conditions \cite{Lyons2017}, a large equilibrium concentration of $V_{\text{Zn}}$ is not expected. Assuming the intermediate formation energy value of 1.77 eV in Table \ref{table:removal-energy} (\textit{n}-type material), an equilibrium concentration of $2\times10^{14}$ to $4\times10^{16}$ cm$^{-3}$ is obtained at 800--1200 $^{\circ}$C. Note that even small changes in the Fermi level position or the chemical potential will change the formation energy of $V_{\text{Zn}}$, and thus the equilibrium concentration, by a significant amount. For instance, intentional \textit{n}-type doping promotes the formation of $V_{\text{Zn}}$ and $V_{\text{Zn}}$-donor complexes in ZnO \cite{Look2011,Demchenko2011,Johansen2015}. For melt grown (MG) ZnO from Cermet inc., and vapor phase (VP) grown ZnO from EaglePicher, positron annihilation spectroscopy (PAS) measurements show a concentration of open-volume defects related to $V_{\text{Zn}}$ below the PAS detection limit of $1\times10^{15}$ cm$^{-3}$ \ \cite{Knutsen2012,Zubiaga2007}. In addition, EPR measurements show that the concentration of isolated $V_{\text{Zn}}$ in MG ZnO from Cermet inc. is below the detection limit of $1\times10^{13}$ cm$^{-3}$ \ \cite{Stehr2014}. This indicates that: (i) very few $V_{\text{Zn}}$'s are formed during materials growth for these techniques, and/or (ii) most $V_{\text{Zn}}$'s are saturated with donors, e.g., $(V_{\text{Zn}}3\text{H})$, and thus not detected using PAS and EPR. The latter interpretation is strongly favored by the data in Table \ref{table:removal-energy}.

\subsection{\label{sec:defect_luminescence}Defect complex luminescence}

Optical transitions involving electron capture from the CB by $(V_{\text{Zn}}\text{Al}_{\text{Zn}})$, $(V_{\text{Zn}}\text{Si}_{\text{Zn}})$, $(V_{\text{Zn}}\text{H})$, $(V_{\text{Zn}}2\text{H})$, $(V_{\text{Zn}}\text{Al}_{\text{Zn}}2\text{H})$ and $(V_{\text{Zn}}\text{Si}_{\text{Zn}}\text{H})$ have been explored. The effective parameters for all transitions are provided in Table \ref{table:effective-parameters}. We find effective normal-mode frequencies $\hbar\Omega_{\text{g/e}}$ between 22 and 33 meV, total mass-weighted distortions between 2.6 and 3.9 amu$^{1/2}$\AA, and Huang-Rhys factors between 24 and 37 for these $V_{\text{Zn}}$-donor complexes. Moreover, we obtain quite consistent relaxation energies for $V_{\text{Zn}}$ and the $V_{\text{Zn}}$-donor complexes, which is reasonable because the atomic movement is roughly the same in every case, i.e., hole capture is always accompanied by a distinct outward relaxation of the O$^{-}$ ion \cite{Frodason2017}.

\begin{table*}[htb!]
	\caption{Effective parameters for the calculated luminescence transitions; total mass-weighted distortion ($\Delta\text{Q}$), ground- and excited-state normal mode frequencies ($\hbar\Omega_{\text{g/e}}$), classical absorption and emission energies ($E_{\text{abs/em}}$), ZPL energy ($E_{\text{ZPL}}$), peak position (PP), full width at half maximum of the luminescence band (FWHM), ground- and excited-state Huang-Rhys factors ($S_{\text{g/e}}$) and classical barrier for nonradiative capture of photoegenerated holes ($E_{\text{b}}$). The minor discrepancy between $E_{\text{em}}$ and PP is caused by the additional energy term $\frac{1}{2}\hbar(\omega_{\text{e}}-\omega_{\text{g}})$ in the quantum treatment within the harmonic approximation, as well as the slight anharmonicity of the normal modes obtained from the hybrid DFT calculations. \label{table:effective-parameters}}
	\begin{ruledtabular}
		\begin{tabular}{lcccccccc}
			Transition & $\Delta\text{Q}$ (amu$^{1/2}$\AA) & $\hbar\Omega_{\text{g/e}}$ (meV) & $E_{\text{abs/em}}$ (eV) & $E_{\text{ZPL}}$ (eV) & PP (eV) & FWHM (eV) & $S_{\text{g/e}}$ & $E_{\text{b}}$ (meV) \\
			\hline
			$(\mathrm{\mathit{V}_{Zn}Al_{Zn})^{0/-}}$ & 2.70 & 33 / 27 & 2.54 / 1.07 & 1.88 & 1.07 & 0.41 & 25 / 24 & 620 \\
			$(\mathrm{\mathit{V}_{Zn}Al_{Zn})^{+/0}}$ & 2.66 & 33 / 27 & 3.14 / 1.69 & 2.48 & 1.65 & 0.43 & 24 / 24 & 50 \\
			$(\mathrm{\mathit{V}_{Zn}Al_{Zn})^{2+/+}}$ & 3.00 & 28 / 24 & 3.70 / 2.23 & 3.03 & 2.21 & 0.39 & 28 / 27 & 310 \\
			$(\mathrm{\mathit{V}_{Zn}Al_{Zn}\text{H})^{+/0}}$ & 3.09 & 28 / 25 & 3.23 / 1.69 & 2.54 & 1.69 & 0.39 & 30 / 28 & 20 \\
			$(\mathrm{\mathit{V}_{Zn}Al_{Zn}\text{H})^{2+/+}}$ & 3.59 & 25 / 23 & 3.91 / 2.25 & 3.14 & 2.22 & 0.38 & 35 / 34 & 90 \\
			$(\mathrm{\mathit{V}_{Zn}Al_{Zn}2H)^{2+/+}}$ & 3.83 & 25 / 23 & 4.15 / 2.37 & 3.31 & 2.28 & 0.39 & 38 / 37 & 170 \\
			$(\mathrm{\mathit{V}_{Zn}Si_{Zn})^{+/0}}$ & 2.70 & 33 / 27 & 3.14 / 1.43 & 2.24 & 1.39 & 0.43 & 25 / 25 & 180 \\
			$(\mathrm{\mathit{V}_{Zn}Si_{Zn})^{2+/+}}$ & 2.67 & 33 / 27 & 3.56 / 2.10 & 2.89 & 2.05 & 0.43 & 25 / 24 & 10 \\
			$(\mathrm{\mathit{V}_{Zn}Si_{Zn}\text{H})^{2+/+}}$ & 2.98 & 30 / 26 & 3.70 / 2.16 & 3.01 & 2.16 & 0.40 & 29 / 27 & 20 \\
			$(\mathrm{\mathit{V}_{Zn}H)^{0/-}}$ & 2.95 & 30 / 25 & 2.79 / 1.30 & 2.12 & 1.30 & 0.39 & 27 / 27 & 240 \\
			$(\mathrm{\mathit{V}_{Zn}H)^{+/0}}$ & 3.14 & 28 / 25 & 3.40 / 1.83 & 2.66 & 1.81 & 0.39 & 31 / 29 & 10 \\
			$(\mathrm{\mathit{V}_{Zn}H)^{2+/+}}$ & 3.63 & 25 / 22 & 4.09 / 2.53 & 3.37 & 2.46 & 0.37 & 34 / 35 & 170 \\
			$(\mathrm{\mathit{V}_{Zn}2H)^{+/0}}$ & 3.42 & 26 / 24 & 3.54 / 1.85 & 2.73 & 1.84 & 0.37 & 34 / 34 & 10 \\
		\end{tabular}
	\end{ruledtabular}
\end{table*}

\begin{figure*}[!htb]
	\includegraphics[width=0.95\textwidth]{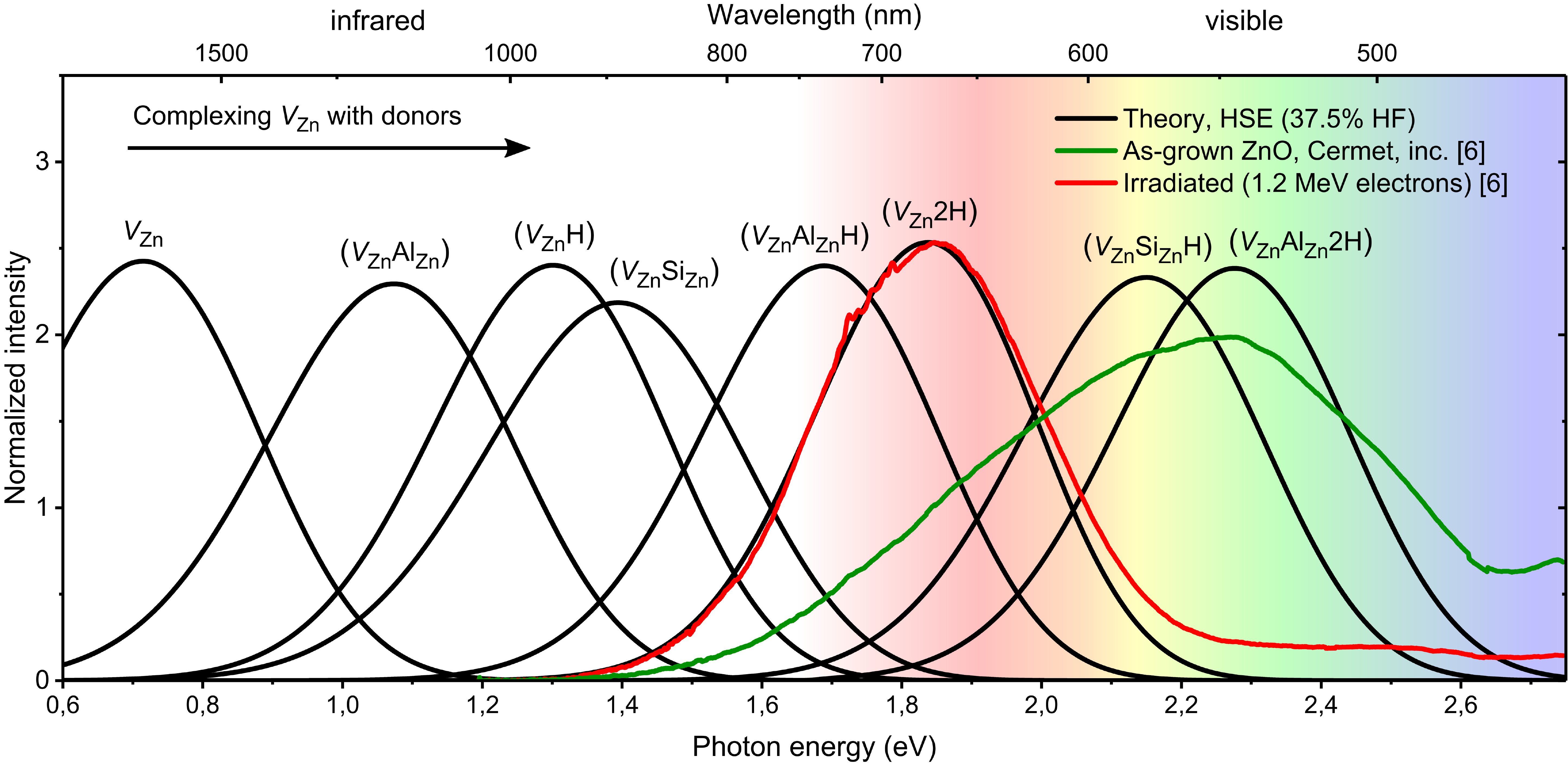}
	\caption{PL spectra showing calculated positions and lineshapes for $V_{\text{Zn}}$-donor complexes. Experimental data from Ref. \onlinecite{Knutsen2012} are included, showing optical emission from MG Cermet inc. samples under 325 nm excitation at 10 K before and after irradiation by 1.2 MeV electrons. The peak intensity of the RL from Ref. \onlinecite{Knutsen2012} is normalized to the calculated ($V_{\text{Zn}}2\text{H}$) peak intensity.}
	\label{fig:luminescence}
\end{figure*}

 We focus on PL experiments carried out at low excitation intensities, and thus restrict the following discussion to transitions involving electron capture by complexes with a single trapped hole polaron. The resulting luminescence lines are shown in Fig. \ref{fig:luminescence}. As indicated in the figure, complexing $V_{\text{Zn}}$ with donors effectively shifts its emission from the infrared region towards the visible range of the spectrum. The main origin of this shift is the electrostatic repulsion between the hole polaron and donor, and its magnitude depends on the nature, configuration and number of donors in the complex. For instance, the emission from ($V_{\text{Zn}}\text{H}$) shifts to a higher energy than ($V_{\text{Zn}}\text{Al}_{\text{Zn}}$), mainly because $U_{\text{Al}}<U_{\text{H}}$ (see Sec. \ref{sec:defect_energetics}). Notably, the calculated luminescence lines for ($V_{\text{Zn}}2\text{H}$), $(V_{\text{Zn}}\text{Al}_{\text{Zn}}2\text{H})$ and $(V_{\text{Zn}}\text{Si}_{\text{Zn}}\text{H})$ are predicted to peak in the visible part of the spectrum. However, the two latter complexes have not been identified experimentally. Moreover, they are positively charged in the ground state, which is likely to have a significant impact on their ability to capture photogenerated holes.
 
During PL experiments carried out at low excitation intensities, photogenerated holes are more likely to be captured by defects with large nonradiative hole capture coefficients like Li$_{\text{Zn}}$ \cite{Alkauskas2014a}. For this reason, the classical barrier for nonradiative hole capture from the VB ($E_{\text{b}}$) is included in Table \ref{table:effective-parameters}. As shown in Fig. \ref{fig:cc_diagram}, this barrier corresponds to the energy required to reach the crossing point between the two potential energy surfaces, i.e, the point where their vibronic coupling is most efficient. Based on the calculated $E_{\text{b}}$ values, one can expect the intensity of the luminescence from, e.g., $(V_{\text{Zn}}2\text{H})$ to be higher than that of $(V_{\text{Zn}}\text{Al}_{\text{Zn}})$, since hole capture is anticipated to take place rapidly for $(V_{\text{Zn}}2\text{H})$. Furthermore, it should be pointed out that the transitions involving electron capture from the CB by $V_{\text{Zn}}^{-}$, $(V_{\text{Zn}}\text{Al}_{\text{Zn}})^{0}$, $(V_{\text{Zn}}\text{H})^{0}$ and $(V_{\text{Zn}}\text{Si}_{\text{Zn}})^{+}$ may have significant nonradiative components; the classical barrier for nonradiative electron capture $E_{\text{A}2}$ (as shown in Fig. \ref{fig:cc_diagram}) is only 0.09 \cite{Frodason2017}, 0.31, 0.42 and 0.60 eV, respectively. These will be lower in the quantum treatment \cite{Alkauskas2014a}, and so the luminescence from an isolated $V_{\text{Zn}}$ \cite{Frodason2017,Nikitenko1992} and $V_{\text{Zn}}$ complexed with a single donor might be weak in reality. Indeed, nonradiative recombination is usually stronger than radiative recombination at low emission energies \cite{Reshchikov2014,Alkauskas2016,Dreyer2016,Nikitenko1992}. Note that nonradiative processes do not interfere with absorption.	
\subsection{\label{sec:red_luminescence}The red luminescence band}

Finally our theoretical predictions are discussed in light of experimental data by Kappers \textit{et al.} \cite{Kappers2008}, Knutsen \textit{et al.} \cite{Knutsen2012} and Vlasenko \textit{et al.} \cite{Vlasenko2005}, where the aforementioned RL emerges after high-energy electron irradiation:

(i) Kappers \textit{et al.} \cite{Kappers2008} carried out a PL and EPR study of VP grown EaglePicher ZnO samples before and after irradiation with 2.0 MeV electrons at RT. After irradiation, the carrier concentration was reduced by three orders of magnitude, and an intense RL band peaking at about 1.8 eV was observed; the donor-bound exciton lines commonly attributed to H ($I_{4}$), Al ($I_{6}$), Ga ($I_{8}$) and In ($I_{9}$) were strongly suppressed, and the EPR signals of Fe$^{3+}$ and $V_{\text{Zn}}^{-}$ were detected in darkness. Photo-EPR data unveiled additional signals from $V_{\text{Zn}}^{0}$, $V_{\text{O}}^{+}$ and $V_{\text{Zn}}$ complexed with H, Al and Ga \cite{Kappers2008,Evans2008}. Based on these observations, the RL was assigned to a DAP transition involving the $V_{\text{Zn}}$ acceptor and shallow donors \cite{Kappers2008}.

(ii) Knutsen \textit{et al}. \cite{Knutsen2012} employed PL, PAS and Hall effect measurements to study MG Cermet inc. ZnO samples irradiated with electrons with energies below and above the threshold for displacement of Zn atoms, as well as samples annealed in Zn-rich and O-rich ambients. The RL band was assigned to a DAP transition involving $V_{\text{Zn}}$-related acceptors and shallow donors. Two PL spectra from Ref. \onlinecite{Knutsen2012}, before and after irradiation by 1.2 MeV electrons, are shown in Fig. \ref{fig:luminescence}. Our calculated luminescence line for the ($V_{\text{Zn}}2\text{H}$) complex is in excellent agreement with the experimental data. This is also the case for $(V_{\text{Zn}}\text{H})^{+/0}$ (Table \ref{table:effective-parameters}), but would require $(V_{\text{Zn}}\text{H})^{-}$ to capture two photogenerated holes in \textit{n}-type material, which is perhaps unlikely for low excitation intensities.

(iii) Vlasenko and Watkins \cite{Vlasenko2005,Vlasenko2005a} performed a PL and optically detected magnetic resonance (ODMR) study of VP grown EaglePicher ZnO samples before and after irradiation with 2.5 MeV electrons, in-situ at 4.2 K, followed by sequential isochronal annealing. The irradiation at 4.2 K produced a broad ``double-humped'' IR band peaking at about 1.38 and 1.65 eV, and suppressed the donor-bound exciton lines. The annealing stages were characterized by the disappearance of the double-humped IR band ($\sim$65-150 K), emergence of the RL band ($\sim$180-230 K), and partial annealing of the RL at RT. This strong evolution below RT suggests migration or rearrangement of point defects, where $\text{Zn}_{i}$ \cite{Janotti2007}, $\text{O}_{i}$ \cite{Huang2009} and $\text{H}_{i}$ \cite{Wardle2006,Johansen2008,Hupfer2017} are the most likely candidates. Indeed, Vlasenko and Watkins \cite{Vlasenko2005} attributed the first annealing stage ($\sim$65-150 K) to $\text{Zn}_{i}$ migration, owing to the disappearance of ODMR signals associated with $\text{Zn}_{i}^{2+}$ and $\text{Zn}_{i}^{2+}$--$V_{\text{Zn}}^{2-}$ Frenkel pairs. No ODMR signal from $\text{O}_{i}$ was observed, but the emergence of the RL band after the second annealing stage was tentatively associated with $\text{O}_{i}$ migration \cite{Vlasenko2005}.

High-energy electron irradiation at cryogenic temperatures generates predominantly elementary point defects $V_{\text{Zn}}^{2-}$, $\text{Zn}_{i}^{2+}$, $V_{\text{O}}^{0}$ and $\text{O}_{i\text{,oct}}^{2-}$ or $\text{O}_{i\text{,split}}^{0}$ \cite{Lyons2017,Bhoodoo2016} (\textit{n}-type material). ZnO is known to exhibit very strong dynamic annealing at RT. Hence, as suggested by Vlasenko and Watkins \cite{Vlasenko2005}, it seems likely that the double-humped IR band originates from close $\text{Zn}_{i}$--$V_{\text{Zn}}$ Frenkel pairs, which annihilate after the first annealing stage. The second annealing stage, characterized by the emergence of the RL, presumably involves migration or rearrangement of $\text{O}_{i}$ \cite{Vlasenko2005} or $\text{H}_{i}$. However, the RL is only observed to increase when the irradiation energy is above the threshold for displacement of Zn atoms \cite{Knutsen2012,Stehr2016}, which strongly suggests that it is related to $V_{\text{Zn}}$. EPR signals from $V_{\text{Zn}}$ complexed with H have been detected by several groups after high-energy electron irradiation at RT \cite{Kappers2008,Evans2008,Son2014}. Migration of $\text{H}_{i}$ and trapping at $V_{\text{Zn}}$, leading to growth of the RL, is consistent with this picture. Moreover, capture of a third H would account for the partial anneal of the RL at RT, which also depends on the electron dose and type of ZnO sample, as observed by Vlasenko and Watkins \cite{Vlasenko2005,Vlasenko2005a}. We underline that further work is required to unravel the microscopic mechanism governing the formation of these complexes \cite{Stehr2014}. Indeed, while H is an omnipresent impurity in ZnO, most of it is expected to occur in a bound state\cite{Wardle2006}, or as ``hidden'' H$_{2}$ molecules \cite{Shi2004}. One could speculate that the $\text{O}_{i}$ generated by the electron irradiation releases free $\text{H}_{i}$ via a reaction with $\text{H}_{\text{O}}$ \cite{Bhoodoo2016}. 

The proposed defect model should also provide an explanation for the unusual thermal quenching of the RL band, starting at about 30 K with a low activation energy of 10-20 meV. Thermal quenching of PL bands caused by DAP transitions in ZnO usually occur through thermal emission of bound holes to the VB, followed by a redistribution of these holes to other recombination channels \cite{Reshchikov2014a}. However, the activation energy for thermal quenching within this model, corresponding to $E_{\text{A}2}$ in Fig. \ref{fig:cc_diagram}, is 0.71 eV for ($V_{\text{Zn}}2\text{H}$), which is too high. Alternatively, the quenching could occur through a gradual conversion from radiative to nonradiative electron capture, but the activation energy within this model, corresponding to $E_{\text{A}1}$ in Fig. \ref{fig:cc_diagram}, is 1.35 eV for ($V_{\text{Zn}}2\text{H}$). 
Knutsen \textit{et al.} \cite{Knutsen2012} suggested that the quenching of the RL might occur via thermal delocalization of excitons bound to shallow donors that take part in the transition. By applying Haynes rule, the ionization energy of the shallow donor was estimated to be very close to the H-related $I_{4}$ line. However, as pointed out by Reshchikov \textit{et al}. \cite{Reshchikov2006}, the efficiency of a DAP recombination is normally limited by the rate of hole capture, meaning that ionization of shallow donors is unlikely to cause the PL quenching, even though we expect hole capture by ($V_{\text{Zn}}2\text{H}$) to be very efficient ($E_{\text{b}}$ is only 10 meV). Knutsen \textit{et al.} \cite{Knutsen2012} also suggested thermal activation of a dominant nonradiative channel, but this model failed to explain why other bands quenched with different activation energies. A detailed study of the RL may be required to understand its thermal quenching behavior. For instance, PL measurements at temperatures above 300 K could be pursued in order to investigate whether a second activation energy, related to thermal emission of bound holes ($E_{\text{A}2}$), can be observed. If ($V_{\text{Zn}}2\text{H}$) is responsible, the activation energy should be approximately 0.7 eV according to our calculations.

\section{\label{sec:conclusion}Conclusion}

Employing hybrid density functional theory calculations, we have shown that $V_{\text{Zn}}$ can form highly stable complexes with the residual shallow donors H, Al, Ga and Si in ZnO. Most of the studied $V_{\text{Zn}}$-donor complexes, and particularly the $(V_{\text{Zn}}n\text{H})$ ones, have a formation energy that is lower than or at least comparable to their individual constituents, meaning that they are likely to form not only during post-growth processing, but even during materials growth. However, the absolute and relative concentration of different complexes is expected to depend strongly on the synthesis method used. Importantly, the results evidence that $V_{\text{Zn}}$-donor complexes can have a crucial impact on the electrical and optical properties of ZnO, and should be considered when interpreting corresponding experimental data. 

Furthermore, by using an effective one-dimensional CC model, defect complex luminescence positions and lineshapes were calculated. The results show that complexing $V_{\text{Zn}}$ with donors effectively shifts the $V_{\text{Zn}}$ emission from the IR towards the visible range of the spectrum. The magnitude of the shift depends on the nature, configuration and number of donors in the complex, and can be interpreted from a polaron energetics model. Based on comparison with a compilation of experimental data in the literature \cite{Kappers2008,Evans2008,Knutsen2012,Vlasenko2005,Wang2009}, the ($V_{\text{Zn}}2\text{H}$) complex is proposed as a potential origin of the high-energy electron irradiation induced RL peaking at about 1.8 eV, which is commonly believed to be $V_{\text{Zn}}$-related \cite{Knutsen2012,Kappers2008,Wang2009}. Above all, our theoretical predictions provide useful input for future optical studies of ZnO, and $V_{\text{Zn}}$-donor complex defect assignment.

\begin{acknowledgments}
	
Financial support was kindly provided by the Research Council of Norway and University of Oslo through the frontier research project FUNDAMeNT (no. 251131, FriPro ToppForsk-program). K.M.J. would like to thank the Research Council of Norway for support to the DYNAZOx project (no. 221992). A.A. was supported by Marie Sk\l{}odowska-Curie Action of the European Union (project \textsc{Nitride}-SRH, Grant No. 657054). The computations were performed on resources provided by UNINETT Sigma2 - the National Infrastructure for High Performance Computing and Data Storage in Norway.
	
\end{acknowledgments}
	
\bibliography{main-text-with-figures}	

\end{document}